# Thermal Conductivity of Chirality-Sorted Carbon Nanotube Networks

Feifei Lian[1], Juan P. Llinas[2], Zuanyi Li[1], David Estrada[2,3] and Eric Pop[1,*]

[1]*Electrical Engineering, Stanford University, Stanford, CA 94305, U.S.A.*
[2]*Electrical & Computer Eng., Univ. Illinois Urbana-Champaign, Urbana, IL 61801, U.S.A.*
[3]*Materials Science & Engineering, Boise State University, Boise, ID 83725, U.S.A.*

The thermal properties of single-walled carbon nanotubes (SWNTs) are of significant interest, yet their dependence on SWNT chirality has been, until now, not explored experimentally. Here we used electrical heating and infrared thermal imaging to simultaneously study thermal and electrical transport in chirality-sorted SWNT networks. We examined solution processed 90% semiconducting, 90% metallic, purified unsorted (66% semiconducting), and as-grown HiPco SWNT films. The thermal conductivities of these films range from 80 to 370 $Wm^{-1}K^{-1}$ but are not controlled by chirality, instead being dependent on the morphology (i.e. mass and junction density, quasi-alignment) of the networks. The upper range of the thermal conductivities measured is comparable to that of the best metals (Cu and Ag) but with over an order of magnitude lower mass density. This study reveals important factors controlling the thermal properties of light-weight chirality-sorted SWNT films, for potential thermal and thermoelectric applications.

[*]Contact: epop@stanford.edu

Carbon nanotube films have a broad range of applications, from solar cells[1,2] and transistors[3] to bolometers[4] and mechanical reinforcement additives for polymers.[5] Recent advances have led to sorting of single-walled carbon nanotubes (SWNTs) into chirally purified (i.e. nearly all-semiconducting or all-metallic) solutions and networks.[6,7] However, previous studies have only focused on the electrical[3] and optical[6] properties of such sorted SWNT films, without reports of their thermal properties, which are also important both fundamentally and practically.[8,9] Individual SWNTs are known to have very high thermal conductivity,[9,10] but the thermal conductivity of SWNT networks and films is typically much lower due to the high thermal resistance of the SWNT junctions.[11,12,13] Nevertheless, the thermal conductivity of SWNT composites could be tuned over nearly four orders of magnitude by changing the alignment of the nanotubes as well as the mass density of the network (and consequently the density of SWNT junctions).[5,14] The ability to tune thermal conductivity in SWNT materials leads to exciting applications for heat spreaders and insulators, as well as potential thermoelectric energy harvesters.[15]

In this work, we simultaneously characterize the electrical and thermal properties of SWNT films with varying fractions of nanotube types (from 90% semiconducting to 90% metallic) by electrical measurements and infrared (IR) thermometry. Using an IR microscope, the real-time temperature profile of SWNT films under electrical bias is mapped. To extract thermal conductivity, a computation model is developed to fit the temperature profile captured by the IR scope, accounting for extrinsic effects such as electrical and thermal contact resistance, which turn out to play key roles. We find that the in-plane thermal conductivity of such solution processed SWNT films ranges from 80–370 $Wm^{-1}K^{-1}$, depending more strongly on SWNT density than on chirality. The high end of these films has thermal conductivity comparable to some of the best metals at room temperature (Ag, Cu) but the SWNT films have ten to twenty times lower mass density.

Figure 1a shows our experimental setup. We use the Quantum Focus Instruments (QFI) InfraScope to measure the temperature of suspended SWNT films at slightly elevated background temperature, $T_0 = 80$ °C, which improves the signal-to-noise ratio.[3,13,16] Suspending the samples across the thermometry platform enables one-dimensional (1D) heat flow and sufficient mechanical support for the suspended film.[5,17] (This is in contrast to our earlier work[3,13] that used much thinner samples on $SiO_2$/Si substrates, where the parasitic heat flow path into the substrate could not be avoided, preventing an analysis of the in-plane thermal conductivity.) The large contacts are electrochemically polished Cu blocks coated with 200 nm/150 nm electron-beam evaporated Ti/Pd, Pd being in contact with the SWNTs. Thin ceramic washers electrically isolate the contacts and control the gap distance ($L$) between the Cu blocks.

We use 90% semiconducting (IsoNanotubes-S), 90% metallic (IsoNanotubes-M), unsorted and purified (PureTubes), and unsorted HiPco SWNTs purchased from NanoIntegris.[18] The IsoNanotubes and PureTubes have SWNT diameters ranging from 1.2–1.7 nm with a mean of 1.4 nm. The metallic tubes have a mean length of 0.5 μm. The semiconducting and purified tubes have a mean length of 1 μm. The HiPco SWNTs have diameters ranging from 0.8–1.2 nm with lengths ranging from 0.1–1 μm. The unsorted HiPco and purified tubes have a semiconducting to metallic ratio of 2:1, i.e. ~33% metallic. We assemble the SWNTs into films on nitrocellulose membranes (MCE MF-Millipore 47 mm diameter, 0.025 μm pores) using vacuum filtration.[19] The filters are dissolved using two 30 minute acetone baths, leaving only the freestanding films. The SWNT films are then suspended across the thermometry platform by directly removing them from the acetone using the measurement platform. As shown in Supplementary[19] Figure S1 the film thicknesses ($t_{\text{film}}$) range from 400–500 nm.

We apply a voltage bias to flow current (in the $y$-direction) through the suspended sample, to induce Joule heating and map the temperature in real time, as shown in Figure 1a,c. The scanning electron microscope (SEM) image in Figure 1b reveals some local alignment and bundling of SWNTs in the network, which we attribute to the vacuum filtration assembly method of the films (additional SEM images in Supplementary[19] Figure S2). Otherwise, the SWNTs are randomly oriented in the ($x$-$y$) plane of the filter, with fewer SWNTs crossing over in the $z$ direction.

Temperature maps like the one in Figure 1c are taken while the device is biased as shown in Figure 1a. The temperature is averaged over a range of pixels in the $x$-direction,[19] across the inner rectangle in Figure 1c. As shown in Figure 1d, the temperature profile peaks in the center of the suspended film with negligible heating at the contacts, indicating good heat sinking by the Pd-coated Cu blocks. We simultaneously obtain electrical measurements of the samples, including the electrical contact resistance using the transfer length method (TLM), by measuring samples with varying suspended separations ($L$ = 0.7–2.0 mm) between the Cu blocks. We combine the thermal imaging maps with a computational model to simultaneously extract the thermal contact resistance and the thermal conductivity from the measured temperature profile. As it turns out, accounting for both electrical and thermal contact resistance is important for obtaining the intrinsic thermal conductivity of the suspended SWNT films.

Figure 2 shows the temperature profiles and SEM images of the (a) semiconducting, (b) unsorted, and (c) metallic films. The top and middle rows show two different biases and dissipated power, respectively, as labeled in the figure insets. The direction of current flow in Figure 2a is from the top to the bottom contact (shown by the arrow), with no measurable change in the temperature profile when reversing the current flow direction. The semiconducting film was the most resistive and therefore had the least heating, largely due to its contact resistance (Supplementary[19] Figure S5). This is not unexpected, because the films are suspended and cannot be gated. Given the voltage biasing scheme, the Joule heating in this film ($\propto V^2/R$) is mainly in the percolation paths that include the less resistive, ~10% metallic SWNTs.[3] The metallic networks have lower electrical resistance and a higher temperature rise for the same applied potential. For the unsorted SWNT films, the temperature rise is in between the metallic and semiconducting films, which is expected since the metallic-semiconducting nanotube junctions have higher electrical resistance and there are an "intermediate" number of metallic percolation paths in this film.[11,20]

To extract the thermal conductivity of the sample, we use a finite element analysis of the 1D heat transfer equation:[21]

$$A\frac{\partial}{\partial y}\left(\kappa\frac{\partial T}{\partial y}\right) + p' - g[T(y) - T_0] = 0 \tag{1}$$

where $A = Wt_{film}$ is the cross-sectional area of the film, $\kappa$ is its in-plane thermal conductivity, $p'$ is the Joule heating power per unit length, $g$ is the heat loss coefficient per unit length to the air or to the contacts (discussed below), $T_0 = 80$ °C is the background temperature of the device, and $T(y)$ is the temperature at location $y$ along the film. This approach implies uniform thermal conductivity and power distribution along the film, which are found to be reasonable assumptions given the uniform density of SWNTs (Figure 1b) and the good fit to the measured data, as we will see below. Since the thermal measurements are done in air, we account for heat loss due to convection and radiation using the heat loss coefficient $g_s$ for one surface of the SWNT film exposed to air:

$$g_s = Wh_{conv} + W\varepsilon\sigma_B\left[T(y)^2 + T_0^2\right]\left[T(y) + T_0\right] \tag{2}$$

where $h_{conv}$ is the heat convection coefficient per unit area[22], $\varepsilon$ is the emissivity of the film as measured by the IR scope (see the Supplementary[19] Figure S4), and $\sigma_B$ is the Stefan-Boltzmann constant. ($h_{conv}$ is taken between 5 to 10 Wm$^{-2}$K$^{-1}$ for natural convection in air[22] and the uncertainty to κ introduced by this range is small, less than 2%, as discussed in Table S2 of the Supplement.[19]) Because the IR scope captures a spatial temperature map of heating in the film, we can use the measured temperature values to calculate $g_s$ at each point "$y$" along the sample to directly calculate the heat loss due to radiation. For the suspended portion of the film, $g = 2g_s$ since both top and bottom surfaces should be taken into account; $p' = (V/R)^2(R - 2R_C)/L$, where $R$ is the measured total electrical resistance of the film and $R_C$ is the electrical contact resistance (described in the Supplement[19] Figure S5 and Table S1). We find that accounting for $R_C$ is essential in such Joule self-heating studies, because excluding it would lead to an overestimation of the power input and corresponding overestimation of the extracted κ, which may have been the case in a previous study.[5] In this work, neglecting $R_C$ would result in an estimated 60% higher κ for the metallic networks.

For the portion of the film supported by the contacts, $p' = 0$ and $g = g_s + Wh_C \approx Wh_C$, where $h_C$ is the thermal contact conductance per unit area between the film and the Pd/Ti/Cu contact. To extract the thermal conductivity of the SWNT film, Eq. (1) is solved by using κ and $h_C$ as fitting parameters for the best fit to the average temperature profile of the film obtained by the IR scope. We verify our results by comparing the 1D model with a three-dimensional (3D) COMSOL thermal model of the SWNT film, shown in Supplementary[19] Figure S3. The uncertainty in the extracted κ due to assumptions about radiation and convection is less than 2%, as discussed in Table S2 of the Supplement.[19] These are smaller than the uncertainty in film thickness due to surface roughness (Supplement[19] Fig. S1), which has between 10-25% effect on the extracted κ values.

Figure 3 shows the thermal model fitted to the temperature profiles of the different SWNT films [averaged along the $x$-direction of the rectangular region in Figure 1(c)]. For the semiconducting, unsorted, and unsorted HiPco films, the model shows excellent agreement with the measurements, validating our assumptions of uniform thermal conductivity and uniform heat generation. For the metallic film, we noticed discrepancies between the model and the experimental data near the contacts. For a better fit, we can slightly increase the gap distance $L$ in the model because the physical length of the suspended SWNT film may be larger than the contact separation (the buckling of metallic films was greater during transfer and suspension, as seen in Figure 2c). Thus we extract a range of thermal contact conductance $h_C = 2 \times 10^3$ to $3.5 \times 10^4$ Wm$^{-2}$K$^{-1}$ for all films, recalling that the contacts are at the ambient temperature $T_0 = 80$ °C. These values are nearly four orders of magnitude lower than those between individual SWNTs[23] or graphene[24] and SiO$_2$, ostensibly due to partial contact between the SWNT network with Pd, due to process and transfer residues, and due to some surface roughness of the metal contacts. The thermal contact conductance of the unsorted films is also at least a factor of two larger than those of the sorted films, which are expected to have some residue from the sorting process (Supplementary[19] Table S1).

In Figure 4, we compare our measured thermal conductivity values with literature values of different carbon nanotube materials, at or near 300 K. Suspended, individual SWNTs[9,24,25] have a very high thermal conductivity near room temperature, ~3000 Wm$^{-1}$K$^{-1}$. A study of aligned multi-wall nanotube (MWNT) films[5] reported the highest in-plane thermal conductivity of such composites to date, ranging from 472–766 Wm$^{-1}$K$^{-1}$. (However, this study[5] did not account for the effects of electrical contact resistance, potentially overestimating the thermal conductivity of the films, as we discussed above.)

The SWNT films in this work have thermal conductivities ranging from approximately 80–370 Wm$^{-1}$K$^{-1}$ when both electrical and thermal contact resistances were carefully taken into account. The highest thermal conductivities were achieved in our purified, unsorted SWNT films, from 117–368 Wm$^{-1}$K$^{-1}$. Our metallic SWNT films have extracted thermal conductivities ranging from 106–137 Wm$^{-1}$K$^{-1}$, which is lower than the sorted semiconducting and the purified, unsorted solution processed films. We attribute the differences to SWNT length (metallic ones being shorter, as stated earlier), possible damage from the sorting process, and the presence of surfactants on the metallic SWNTs. The as-grown HiPco SWNT films have the lowest thermal conductivities ranging from 81–97 Wm$^{-1}$K$^{-1}$. The semiconducting SWNT film thermal conductivities range from 174–220 Wm$^{-1}$K$^{-1}$. The ranges of these measurements correspond to values measured across multiple samples.

Using the Wiedemann-Franz Law, we estimate the electronic contribution to thermal conductivity to be $\kappa_e < 1.1$ Wm$^{-1}$K$^{-1}$ in all our SWNT films (Supplementary[19] Table S1). Thus, we find that the thermal conductivity has essentially no dependence on the chirality or electronic type of the SWNTs, confirming that heat flow is predominantly carried by lattice vibrations (phonons) rather than electrons and that the phonon dispersion changes very little between SWNTs of different chirality.[26, 27] Instead, our results are consistent with the view that the thermal conductivity of SWNT films depends more strongly on the SWNT junctions and the mass density of the films (which also controls the junctions and the SWNT segment lengths between junctions[28]). Previously reported solution-processed SWNTs[29,30] found cross-plane thermal conductivity around 1.68 Wm$^{-1}$K$^{-1}$ for millimeter-thick SWNT films[29] and 2.24 Wm$^{-1}$K$^{-1}$ for MWNT films[30] with mass densities around 0.47 g/cm$^3$. (The cross-plane thermal conductivity is expected to be lower due to the layering of SWNTs during the assembly process.) The mass densities of the quasi-aligned MWNT film study[5] were greater than 1 g/cm$^3$. Our SWNT films had mass densities ranging from 0.5 – 1.1 g/cm$^3$ (Table S1 in the Supplement[19]). In comparison with the thermal conductivities of dry SWNT beds[14] that have thermal conductivities ranging from 0.13 to 0.19 Wm$^{-1}$K$^{-1}$ (with mass density 0.2 to 0.45 g/cm$^3$), the solution-processed films studied here are more thermally conductive in the in-plane direction. This can be attributed to many factors such as the higher mass density of our films, the length of the SWNTs, bundling of the SWNTs, and the intrinsic thermal conductivity of individual nanotubes within the network.

Our experimental findings are consistent with theoretical values predicted by Volkov and Zhigilei,[31] who explored the strong influence of the mass density, length and thermal conductivity of individual SWNTs on the network thermal conductivity. In this context, part of the difference in thermal conductivities between the various nanotube films in our study may be due to different intrinsic $\kappa$ of the SWNTs in the films. For example, it is known that the effective $\kappa$ for both SWNTs and graphene depends on their length when it is comparable to the phonon mean free path.[25,32] The metallic SWNTs are shorter (~0.5 μm) and potentially more damaged than the semiconducting or purified SWNTs (~1 μm) after the sorting process, which is consistent with the observed lower overall $\kappa$ for the metallic SWNTs films.

In summary, we used a combination of IR thermometry and electrical measurements to characterize solution-processed films with controlled density of metallic and semiconducting SWNTs. Metallic films have higher electrical conductivity than semiconducting films (as expected) but lower thermal conductivity due to shorter tube lengths, which also leads to greater SWNT junction density. More importantly, the thermal conductivity of solution-processed SWNT networks is higher than that of dry-assembled SWNT beds[14] due to the vacuum filtration assembly process. Overall, we find that chirality plays essentially no role on thermal, which are controlled by the individual SWNT lengths, and overall junction and mass density of the SWNTs.

From a metrology standpoint, this study highlights the importance of adjusting for electrical and thermal contact resistance in measurements on such suspended films, before intrinsic thermal parameters can be the deduced accurately. From a practical standpoint, these are important findings for lightweight heat spreaders and for thermoelectric energy harvesters. In particular, for thermoelectric applications[15] our results underscore that the figure of merit (ZT) of a SWNT sample cannot be estimated based on previously measured results on different samples.[14] Rather, the thermal conductivity of SWNT thermoelectrics must be measured independently, because these quantities are sensitive to the morphology of the sample.

This work was supported in part by the Presidential Early Career (PECASE) Grant W911NF-13-1-0471 through the Army Research Office, the National Science Foundation (NSF) Grant 13-46858, and the NSF Center for Power Optimization of Electro-Thermal Systems (POETS). We are indebted to A.D. Liao and J.D. Wood for helpful discussions and technical support.

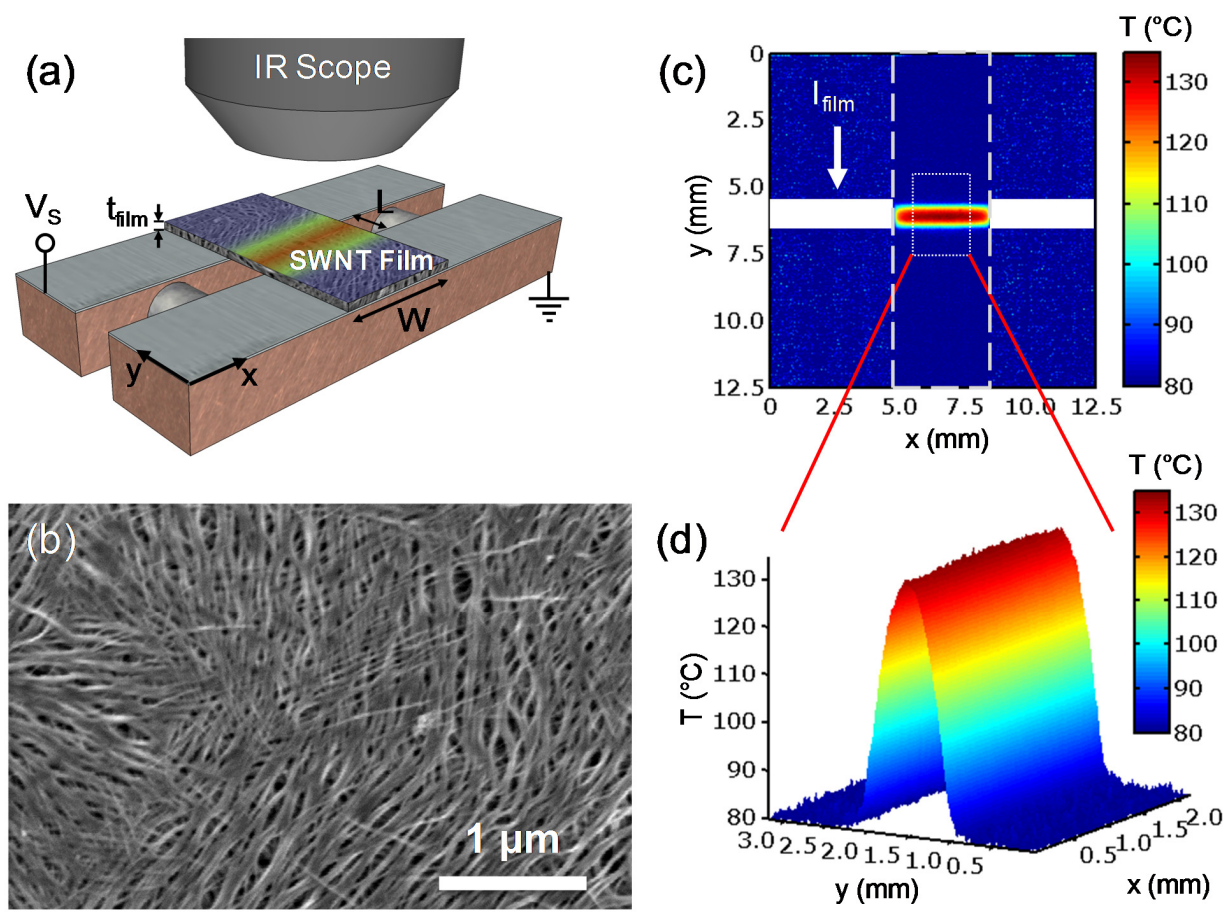

**Figure 1.** (a) Schematic of the thermometry platform and the experimental set up. SWNT films are suspended across two Pd-coated Cu blocks that are electrically isolated by ceramic washers. (b) SEM image of the SWNT film after vacuum filtration. The SWNTs are bundled and randomly in-plane oriented. (c) Temperature map of the SWNT film across the metal contacts. White dashed lines show the edges of the SWNT film, and current flows in the direction of the arrow. (d) The zoomed-in temperature profile of the suspended SWNT film across the gap. The 1D temperature profiles in Figure 3 are averaged along the *x*-direction of such maps.

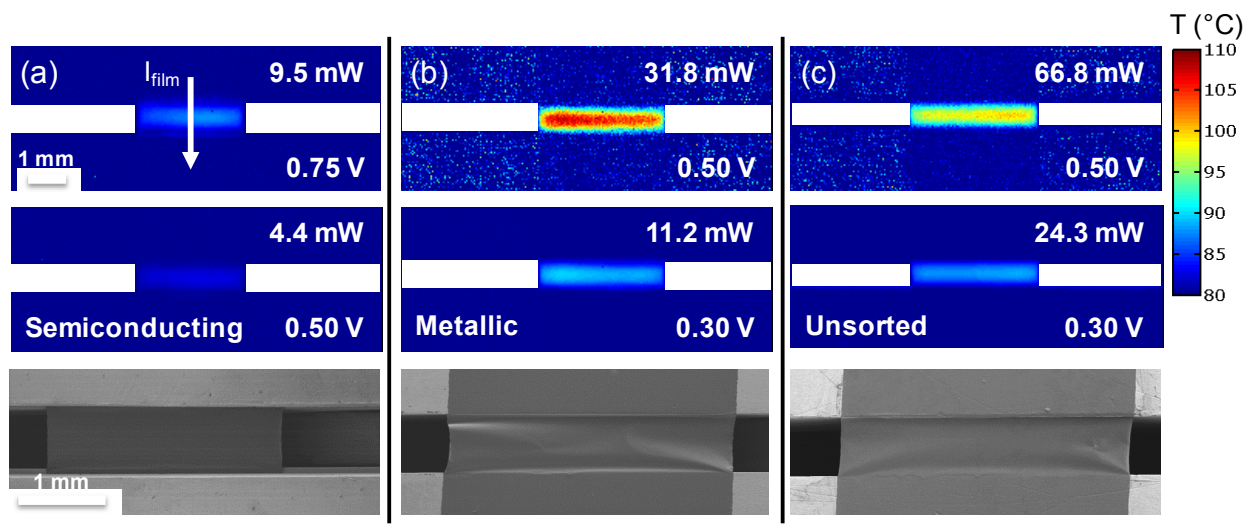

**Figure 2.** Temperature profiles and SEMs of (a) semiconducting, (b) unsorted, and (c) metallic SWNT. Top and middle panels correspond to higher and lower power applied to the networks, respectively. The insets list the applied voltages and the power dissipated in the suspended portion of the films, excluding contact resistance, $(V/R)^2(R – 2R_C)$. The vertical arrow shows the current flow direction. Some bowing in the films from the transfer process can be seen in the SEMs for the metallic and unsorted networks.

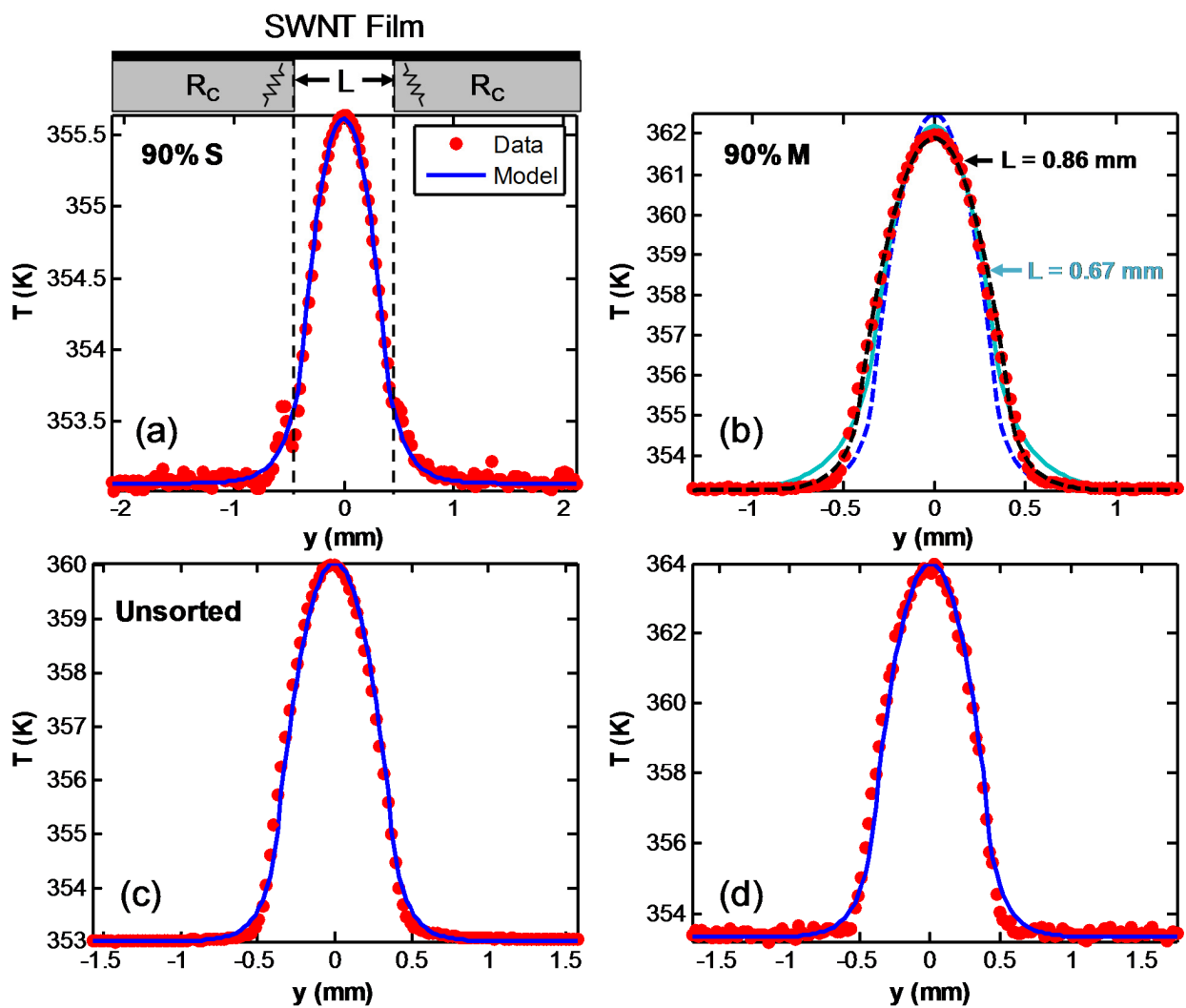

**Figure 3.** Averaged temperature profiles (symbols) fitted by the model (lines) for (a) the semiconducting film, (b) metallic film, (c) purified unsorted film, and (d) as-grown HiPco film. The upper panel in (a) illustrates the role of the electrical and thermal contact resistance. In (b), there is a slight discrepancy between the model and the measured temperature profile for the metallic film. The light blue dashed line shows the model using the measured gap distance ($L = 0.67$ mm) as the length of the suspended portion of the film. The black dashed line denotes the model adjusted using a larger gap distance ($L = 0.86$ mm). The blue dashed line shows the effect of fixing the thermal contact conductance while using the physical gap distance.

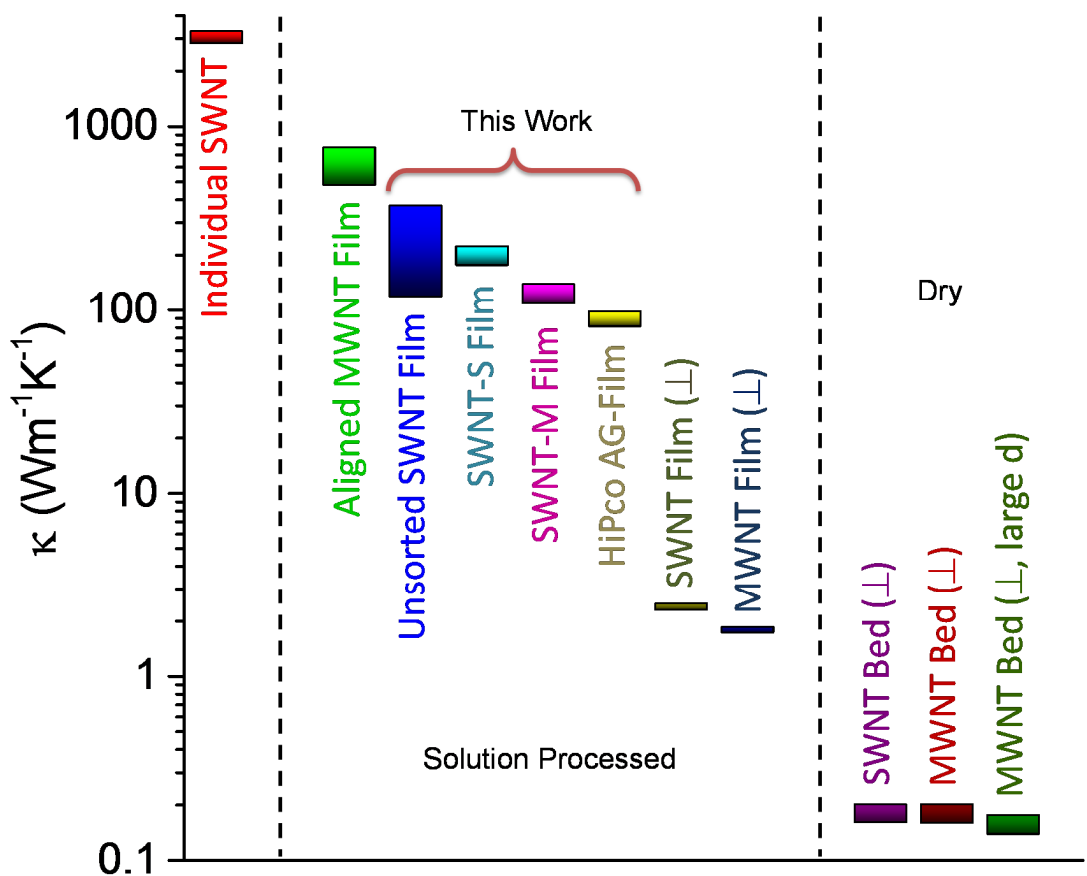

**Figure 4.** Summary of thermal conductivities of carbon nanotube films and composites near room temperature, including the results of this work: unsorted SWNT films, 90% semiconducting (SWNT-S) films, 90% metallic (SWNT-M) films, and HiPco as-grown (AG) films. The thermal conductivities of aligned MWNT films,[5] thick SWNT[29] and MWNT films,[30] and SWNT and MWNT dry beds[14] are also shown for comparison. ($\perp$) denotes cross-plane thermal conductivity from their respective references. The SWNT composites are separated into solution-processed films and dry-assembled mats; large diameter (*d*) mats have diameters ranging from 60-100 nm.
13

# Supplementary Information

# Thermal Conductivity of Chirality-Sorted Carbon Nanotube Networks

Feifei Lian[1], Juan P. Llinas[2], Zuanyi Li[1], David Estrada[2,3] and Eric Pop[1,*]

[1]*Electrical Engineering, Stanford University, Stanford, CA 94305, U.S.A.*
[2]*Electrical & Computer Eng., Univ. Illinois Urbana-Champaign, Urbana, IL 61801, U.S.A.*
[3]*Materials Science & Engineering, Boise State University, Boise, ID 83725, U.S.A.*

[*]Contact: epop@stanford.edu

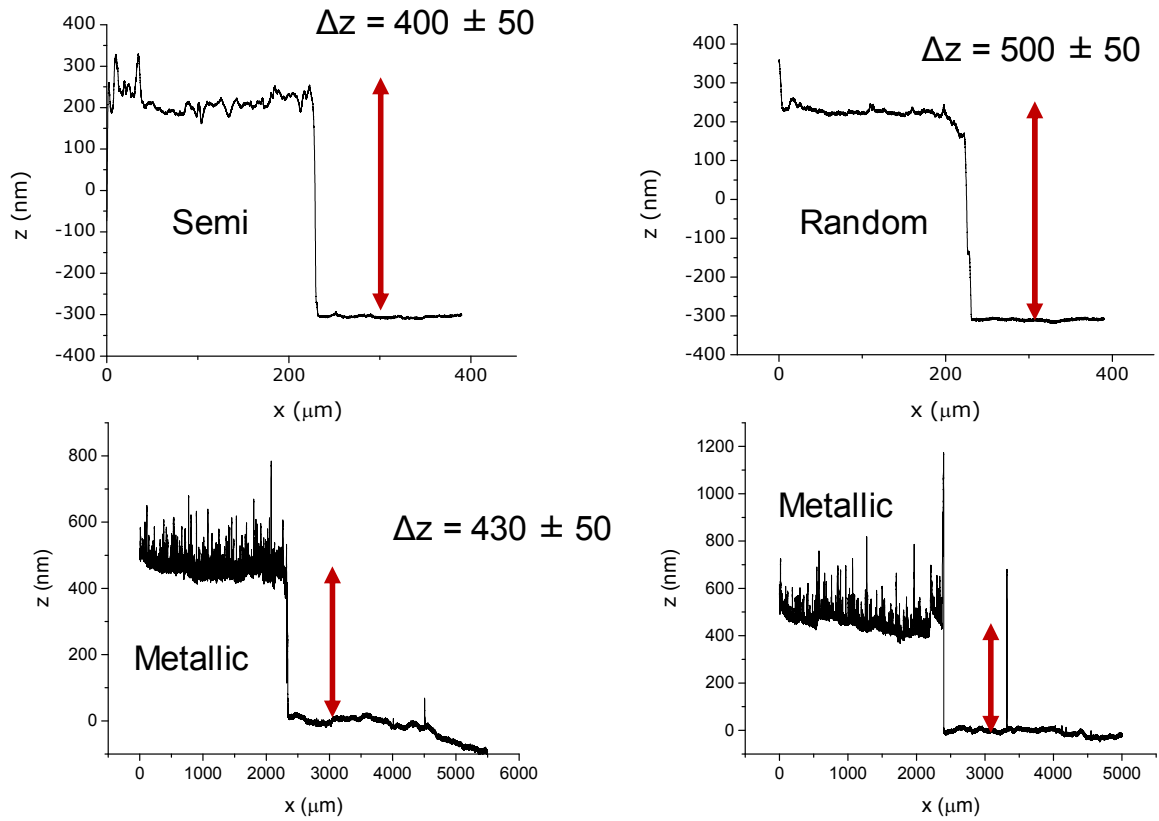

**FIG. S1.** Profilometer measurements of the SWNT films on Si substrates. Thicknesses ($t_{film}$) were used for extracting thermal conductivity in the model. Metallic films had more surface roughness, but their overall film thickness is estimated to be 400-500 nm. The mass density of the film is also calculated from the thickness and the mass of the SWNTs used in the film assembly.

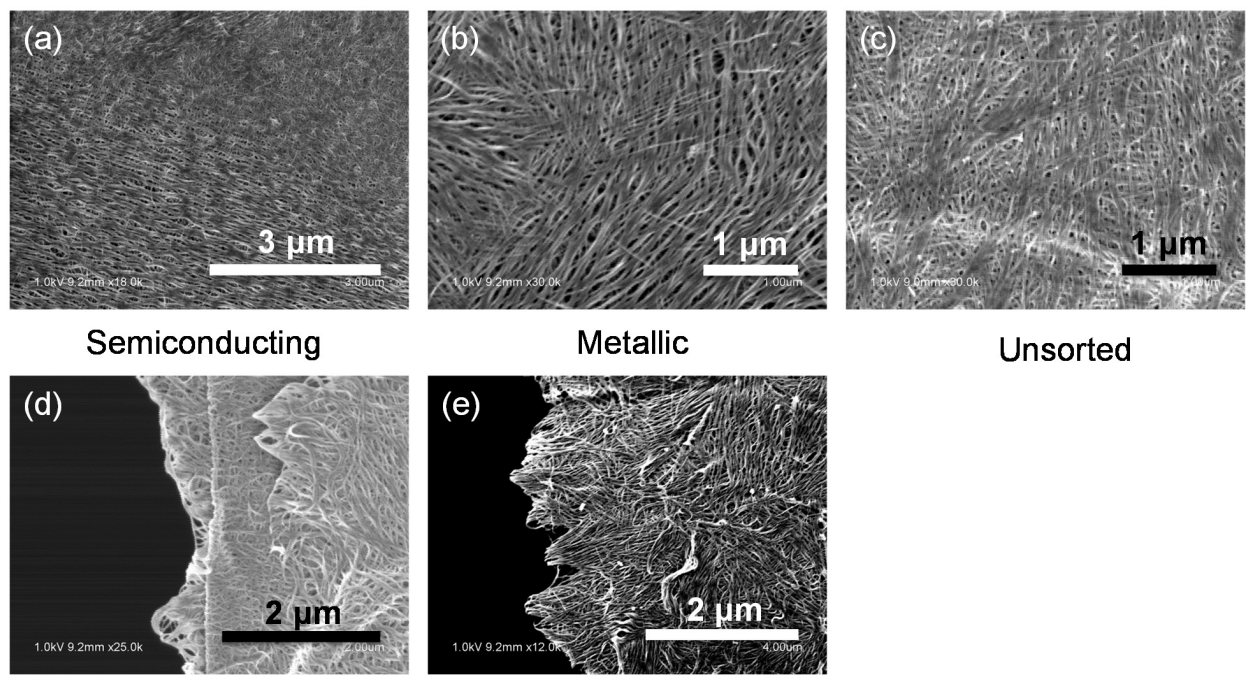

**FIG. S2.** Scanning electron microscope (SEM) images of our solution-assembled SWNT networks. (a) – (c) show SEMs of the center of the film for semiconducting, metallic, and unsorted networks, respectively. Figures (d) and (e) show the edge of the film for the semiconducting and metallic networks respectively. The edge roughness of the film is due to the cutting of the film following the vacuum filtration assembly.

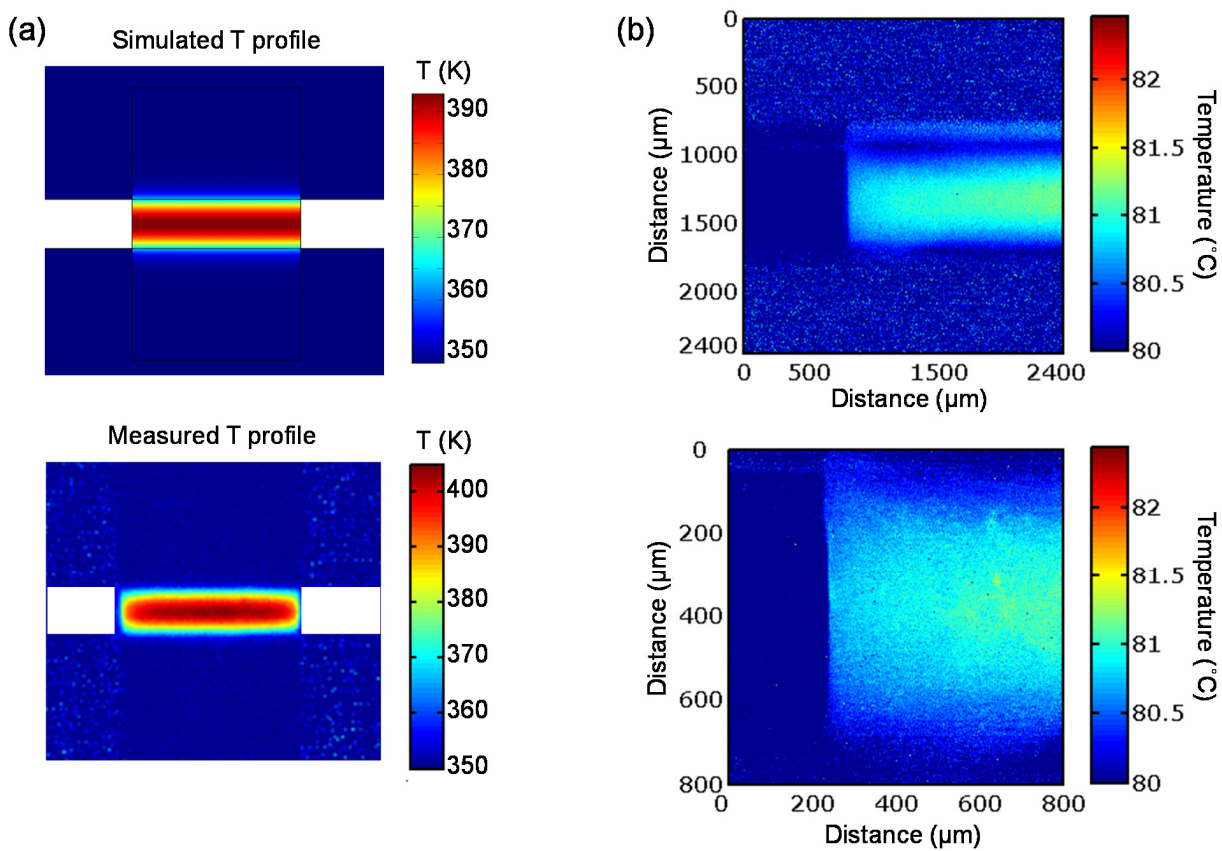

**FIG. S3.** (a) Shows a comparison of a 3D finite-element (COMSOL) model (top) with the measured temperature profile at 5× magnification (bottom). Non-uniformities in the measured profile are an artifact of the IR scope due to the spatial resolution of the objective. Figure (b) shows the temperature profile of the films at 10× and 15× magnification.

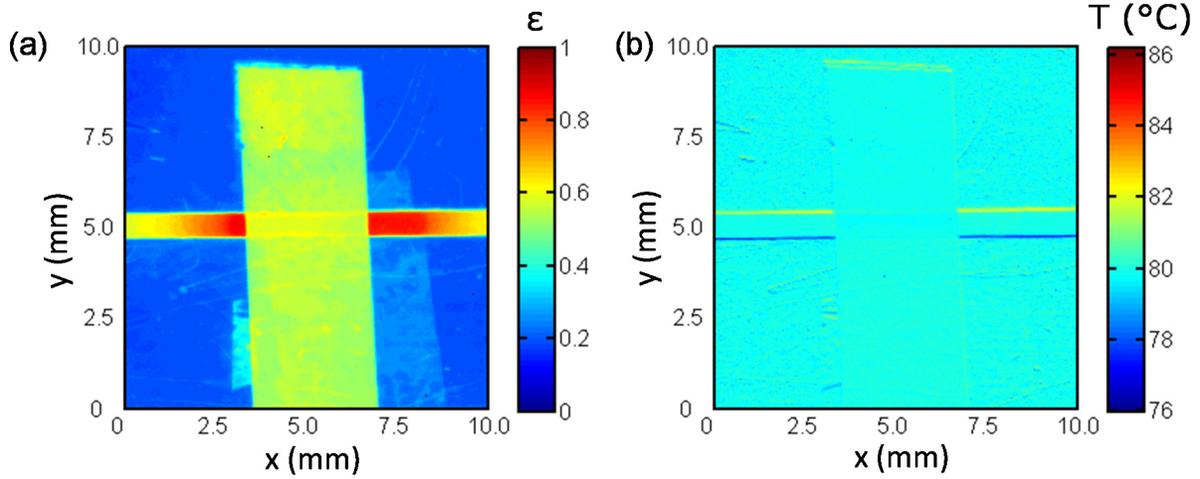

**FIG. S4.** (a) Reference radiance measurement of Unsorted film from IR scope. Values for emissivity ($\varepsilon$) are used to calculate the radiative heat loss from the SWNT film. The average emissivity of the SWNT films are listed in Table S2. We directly measured the emissivity of the Pd coated Cu contacts to be $\varepsilon \approx 0.16$, as expected. (b) Background temperature measurement performed without any applied bias across the SWNT film. Slight color difference at edges of metal contacts are due to reflection from the edges of the contacts.

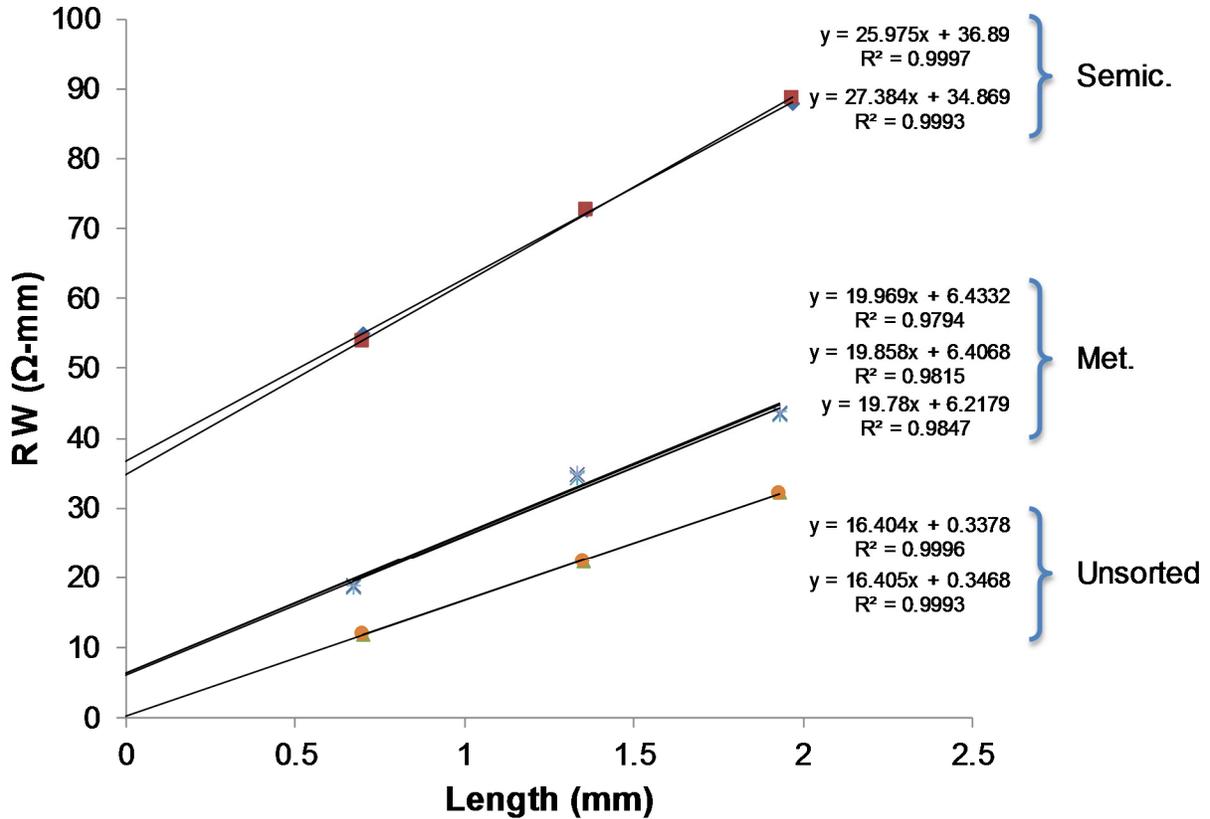

**FIG. S5.** Transfer length method (TLM) plot of electrical resistances of the films (multiplied by their width) as a function of suspended film length. Symbols are experimental data and lines are linear fits. The vertical intercept represents twice the electrical contact resistance ($2R_CW$), the slope represents the SWNT film sheet resistance ($R_{sh}$), and the horizontal intercept represents an estimate of the transfer length ($2L_T$). Multiple measurements were taken at several voltage biases, and the equation of the linear fit to each is given in the inset. The unsorted samples were biased at 0.5 V and 0.3 V, and the metallic samples were biased at 0.75 V, 0.5 V, and 0.3 V with no noticeable change in resistance of either sample. The semiconducting sample was biased at 0.75 V and 0.5 V and a slight decrease in resistance was observed at the higher bias.

Contact resistance is most significant for the semiconducting network and varies depending on the surface roughness of the contacts, as well as the presence of residue between the SWNT film and the metal surface. (We measured the RMS roughness of the metal contacts to be ~165 nm.) We believe the unsorted networks have lower contact resistance since they have much less damage (no sorting) and much less residue. Importantly, the electrical contact resistance was always taken into account in all power input calculations for the extraction of thermal conductivity.

| Film Type | σ (S/m) | $R_{sh}$ (Ω/□) | $2R_CW$ (Ω·mm) | $L_T$ (mm) | $\kappa_e$ (Wm$^{-1}$K$^{-1}$) | $t_{film}$ (nm) | Mass (μg) | ρ (g/cm³) | $h_c$ (m²K/W) |
|---|---|---|---|---|---|---|---|---|---|
| 90%-S | ~ 8.34×10⁴ | ~26 | ~35.0 | ~0.65 | ~0.74 | 450 ± 50 | 400 | ~0.51 | 3.0×10³ – 1.6×10⁴ |
| 90%-M | ~ 1.17 ×10⁵ | ~20 | ~6.35 | ~0.16 | ~1.07 | 430 ± 50 | 800 | ~1.07 | 6.0×10³ – 8.0×10³ |
| Unsorted | ~ 1.22 ×10⁵ | ~16 | ~0.34 | ~0.10 | ~1.10 | 500 ± 50 | 500 | ~0.58 | 1.5×10⁴ – 2.0×10⁴ |

**Table S1.** Electrical and physical properties for the 90% semiconducting, 90% metallic, and purified unsorted films. Electronic contribution to thermal conductivity is estimated using the Wiedemann-Franz Law, $\kappa_e = \sigma L_0 T$ where σ is the electrical conductivity extracted from the TLM measurements, $L_0$ is the Lorenz constant and T is the temperature. The unsorted, purified tubes are higher quality than the sorted semiconducting and metallic networks, leading to the higher electrical conductivity. The mass density of the metallic network is also twice as high as the semiconducting and the unsorted networks, which leads to higher junction density. We believe the higher junction density and shorter SWNT lengths (also indicative of more damage) are responsible for the thermal conductivity of the metallic networks being somewhat lower.

| | κ (Wm$^{-1}$K$^{-1}$) | parameter | $t_{film}$ (nm) | W (mm) | $2R_c$ (Ω) | $h_{conv}$ (Wm$^{-2}$K$^{-1}$) | ε |
|---|---|---|---|---|---|---|---|
| 90%-S | 174 – 220 | input error | 400±50 | 2.5±0.1 | 35.9±1.0 | 5±5 (i.e. 0–10) | 0.58±0.01 |
| | | κ uncertainty | ~12% | ~4% | ~3% | ~2% | ~0.1% |
| 90%-M | 107 – 137 | input error | 430±50 | 3.6±0.05 | 6.35±0.1 | 5±5 (i.e. 0–10) | 0.37±0.01 |
| | | κ uncertainty | ~25% | ~1% | ~1% | ~2% | ~0.1% |
| Unsorted | 286 – 368 | input error | 500±50 | 3.3±0.03 | 0.34±0.005 | 5±5 (i.e. 0–10) | 0.57±0.01 |
| | | κ uncertainty | ~10% | ~1% | ~1% | ~2% | ~0.1% |

**Table S2.** Calculated uncertainty analysis for the extracted SWNT film thermal conductivity (κ) in our measurements. We consider the errors from the film thickness, contact resistance, and convection and radiation losses. The main uncertainty resulted from the thickness of the film which can be seen from Fig. S1. We note that the true cross-sectional area of the SWNT film is *not* $Wt_{film}$ because the SWNTs are not fully packing the rectangular parallelepiped with volume $WLt_{film}$ (see Fig. 1). We can estimate the fill factor by two means: 1) the estimated mass density is ~1.1 g/cm³ which is approximately 50% that of graphite, indicating about 45% fill factor in the network. 2) the estimated thermal κ is approximately 10% that of graphite. We regard the former estimate as more accurate for the fill factor, and attribute the thermal κ being lower than $0.45\kappa_{graphite}$ to the effects of intertube junctions and misalignment.